\documentclass[aps,prl,longbibliography,twocolumn,superscriptaddress,floatfix]{revtex4-1}
\usepackage{amsmath}
\usepackage{amssymb}
\usepackage{amsthm}
\usepackage{amsfonts}
\usepackage{listings}
\usepackage{enumerate}
\usepackage{latexsym}
\usepackage{psfrag}
\usepackage{bm}
\usepackage[all]{xy}
\usepackage{graphicx}
\usepackage{subfigure}
\usepackage[pdftex,colorlinks=false]{hyperref}
\usepackage{color}
\usepackage{mathtools}
\usepackage{eufrak}
\usepackage{ gensymb }
\usepackage{lipsum}
\begin{document}

\title{Gapped spin-1/2 spinon excitations in a new kagome quantum spin liquid compound Cu$_3$Zn(OH)$_6$FBr}
\author{Zili Feng / 冯子力}
\thanks{These two authors contribute to this work equally.}
\affiliation{Institute of Physics, Chinese Academy of Sciences, Beijing 100190, China}
\author{Zheng Li}
\thanks{These two authors contribute to this work equally.}
\affiliation{Institute of Physics, Chinese Academy of Sciences, Beijing 100190, China}
\affiliation{School of Physical Sciences, University of Chinese Academy of Sciences, Beijing 100190, China}
\author{Xin Meng}
\affiliation{Institute of Physics, Chinese Academy of Sciences, Beijing 100190, China}
\author{Wei Yi}
\affiliation{Institute of Physics, Chinese Academy of Sciences, Beijing 100190, China}
\author{Yuan Wei}
\affiliation{Institute of Physics, Chinese Academy of Sciences, Beijing 100190, China}
\author{Jun Zhang}
\affiliation{State Key Laboratory of Surface Physics, Department of Physics, and Laboratory of Advanced Materials, Fudan University, Shanghai 200433, China}
\author{Yan-Cheng Wang}
\affiliation{Institute of Physics, Chinese Academy of Sciences, Beijing 100190, China}
\author{Wei Jiang}
\affiliation{Department of Materials Science and Engineering, University of Utah, Salt Lake City, Utah 84112, USA}
\author{Zheng Liu}
\affiliation{Institute for Advanced Study, Tsinghua University, Beijing 100084, China}
\author{Shiyan Li}
\affiliation{State Key Laboratory of Surface Physics, Department of Physics, and Laboratory of Advanced Materials, Fudan University, Shanghai 200433, China}
\affiliation{Collaborative Innovation Center of Advanced Microstructures, Nanjing 210093, China}
\author{Feng Liu}
\affiliation{Department of Materials Science and Engineering, University of Utah, Salt Lake City, Utah 84112, USA}
\author{Jianlin Luo}
\affiliation{Institute of Physics, Chinese Academy of Sciences, Beijing 100190, China}
\affiliation{School of Physical Sciences, University of Chinese Academy of Sciences, Beijing 100190, China}
\affiliation{Collaborative Innovation Center of Quantum Matter, Beijing 100190, China}
\author{Shiliang Li}
\affiliation{Institute of Physics, Chinese Academy of Sciences, Beijing 100190, China}
\affiliation{School of Physical Sciences, University of Chinese Academy of Sciences, Beijing 100190, China}
\affiliation{Collaborative Innovation Center of Quantum Matter, Beijing 100190, China}
\author{Guo-qing Zheng}
\email{gqzheng@iphy.ac.cn}
\affiliation{Institute of Physics, Chinese Academy of Sciences, Beijing 100190, China}
\affiliation{Department of Physics, Okayama University, Okayama 700-8530, Japan}
\author{Zi Yang Meng}
\email{zymeng@iphy.ac.cn}
\affiliation{Institute of Physics, Chinese Academy of Sciences, Beijing 100190, China}
\author{Jia-Wei Mei}
\email{jiawei.mei@gmail.com}
\affiliation{Institute for Quantum Science and Engineering, and Department of Physics, Southern University of Science and Technology, Shenzhen 518055, China}
\affiliation{Department of Materials Science and Engineering, University of Utah, Salt Lake City, Utah 84112, USA}
\affiliation{Beijing Computational Science Research Center, Beijing 100193, China}
\author{Youguo Shi}
\email{ygshi@iphy.ac.cn}
\affiliation{Institute of Physics, Chinese Academy of Sciences, Beijing 100190, China}

\date{\today}
\begin{abstract}
We report a new kagome quantum spin liquid candidate Cu$_3$Zn(OH)$_6$FBr, which does not experience any phase transition down to 50 mK, more than three orders lower than the antiferromagnetic Curie-Weiss temperature ($\sim$ 200 K). A clear gap opening at low temperature is observed in the uniform spin susceptibility obtained from $^{19}$F nuclear magnetic resonance measurements. We observe the characteristic magnetic field dependence of the gap as expected for fractionalized spin-1/2 spinon excitations. Our experimental results provide firm evidence for spin fractionalization in a topologically ordered spin system, resembling charge fractionalization in the fractional quantum Hall state. 
\end{abstract}

\maketitle
When subject to strong geometric frustrations, quantum spin systems may achieve paramagnetic ground states dubbed quantum spin liquid (QSL)~\cite{Anderson1987}. It is characterized by the pattern of long-range quantum entanglement that has no classical counterpart~\cite{Wen2004,Kitaev2006,Levin2006}. QSL is an unambiguous Mott insulator whose charge gap is not associated with any symmetry breaking~\cite{Anderson1987}. It is related to the mechanism of high-temperature superconductivity~\cite{Anderson1987} and the implementation of topological quantum computation~\cite{Kitaev2003}. The underlying principle of QSL, i.e. topological orders due to quantum entanglement~\cite{Kitaev2006,Levin2006}, is beyond the Landau symmetry-breaking paradigm~\cite{Wen2004} and has been realized in fractional quantum Hall systems~\cite{Tsui1982}, resulting in fractionalized $e/3$ charged anyon~\cite{Laughlin1983,Picciotto1997}. Similarly, fractionalized spin-1/2 spinon excitations are allowed in QSL~\cite{Kivelson1987,Read1989,Read1991,Wen1991}.

Kagome Heisenberg antiferromagnets are promising systems for the pursuit of QSL~\cite{Lee2008,Balents2010,Norman2016}. For example, herbertsmithite ZnCu$_3$(OH)$_6$Cl$_2$ is a famous kagome system, which displays a number of well-established QSL behaviors~\cite{Shores2005,Helton2007,Mendels2007,Zorko2008,Imai2008,
Vries2009,Helton2010,Imai2011,Jeong2011,Han2012,Han2012a,
Han2011,Han2012,Imai2008,Olariu2008,Fu2015}. Inelastic neutron scattering measurements have detected continuum of spin excitations~\cite{Han2012} while nuclear magnetic resonance (NMR) measurements suggest a finite gap at low temperature~\cite{Fu2015}. However, multiple NMR lines of nuclear spins with $I>1/2$ can not be easily resolved, particularly in the presence of residual interkagome Cu$^{2+}$ spin moments even in high-quality single crystals~\cite{Imai2008,Olariu2008,Imai2011,Fu2015}. Furthermore, although it is commonly accepted that the quantum number of spinons is spin-1/2, no direct evidence has been observed~\cite{Fu2015}. Therefore, it is crucial to find new QSL systems to unambiguously demonstrate the spin-1/2 quantum number of spinons.

Recently, barlowite Cu$_4$(OH)$_6$FBr has attracted much attention as a new kagome system with minimum disorder~\cite{Elliot2010,Elliot2014,Han2014,Jeschke2015,Liu2015,Han2016}. As opposed to herbertsmithite with $ABC$-stacked kagome planes, barlowite crystallizes in high-symmetry hexagonal rods owing to direct $AA$ kagome stacking. 
It has also been found that the in-plane Dzyaloshinskii-Moriya interaction in barlowite is an order of magnitude smaller than that in herbertsmithite~\cite{Han2016}. 
Consequently, the QSL physics has been suggested to be present at relative high temperature. Unfortunately, the material goes through an antiferromagnetic transition at $\sim15$~K~\cite{Han2014,Han2016}. It has thus been proposed that substituting the interkagome Cu$^{2+}$ sites with non-magnetic ions may suppress the magnetic transition and ultimately lead to a QSL ground state~\cite{Han2014,Liu2015,Guterding2016,Norman2016}.

\begin{figure*}
\centering
\includegraphics[width=1.8\columnwidth]{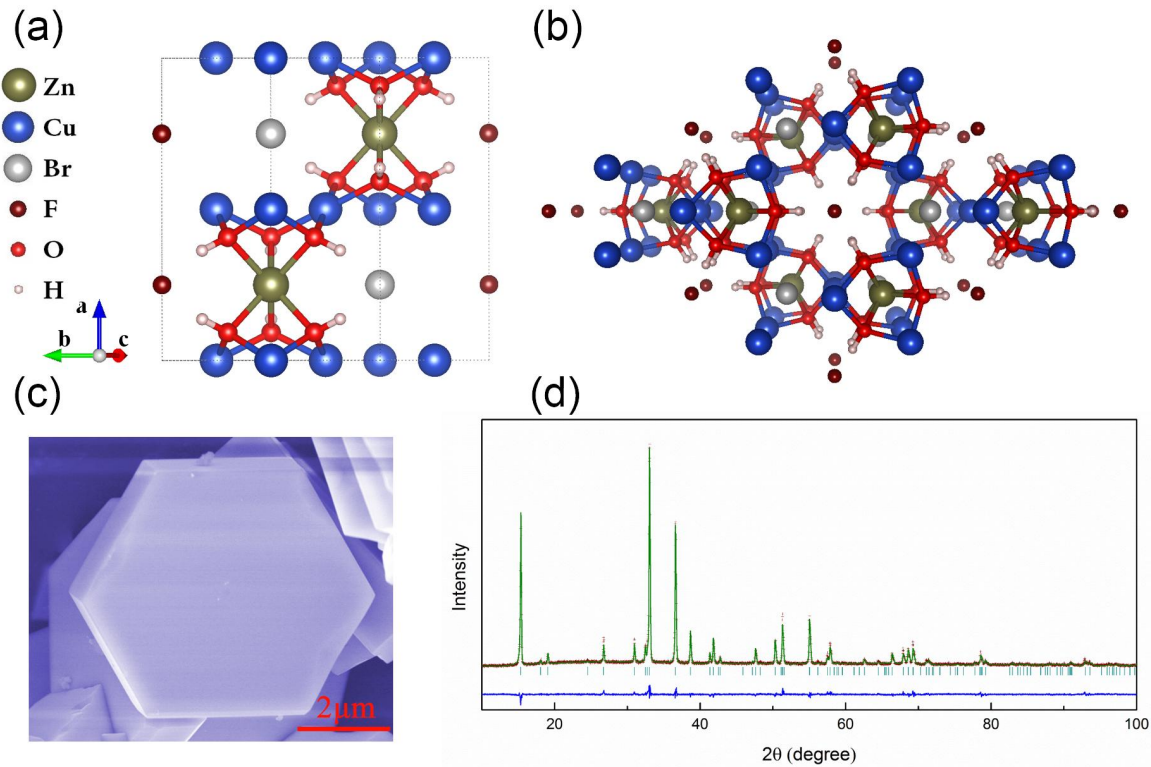}
\caption{(a) Schematic crystal structure of Cu$_3$Zn(OH)$_6$FBr with copper Cu$^{2+}$ ions (blue) forming the kagome planes $AA$ stacked along $c$-axis. Kagome planes are separated by non-magnetic Zn$^{2+}$ (blond) ions. (b) Top view of the Cu$_3$Zn(OH)$_6$FBr crystal structure, where F (brown) is in the center between two hexagons of two kagome Cu planes. (c) Scanning electron microscope image of crystal grain in the polycrystalline samples. (d) Measured (brown ``+'') and calculated (green line) XRD diffraction intensities of polycrystalline samples. The blue curve indicates the difference between the measured and calculated intensities. The vertical lines indicate peak positions. }
\label{fig:figure1}
\end{figure*}

\begin{table}[b]
\centering
\caption{Structure parameters of Cu$_3$Zn(OH)$_6$FBr at room temperature. Space group $P6_3/mmc$; $a = b = 6.6678(2)$ \AA, $c = 9.3079(3)$ \AA.}
\label{tab:XRD}
\begin{tabular}{|cccccc|}
\hline
Site & $w$ & $x$ & $y$ & $z$ & $B$ (\AA$^2$)  \\ \hline
Cu & $6g$ &0.5&0&0&1.48(6)  \\
Zn & $2d$ &1/3 &2/3 &3/4 &1.93(8) \\
Br & $2c$ &2/3 &1/3 &3/4 &1.99(5)\\ 
F  & $4b$ &0.0 &0.0 &3/4 &0.34(2)\\
O  & $12k$ &0.1887~ &0.8113(5) &0.9021(7) &2.22(2)\\
H  & $12k$ &0.1225 &0.8775 &0.871 &1.0\\\hline
\end{tabular}
\end{table}
In this Letter, we report a new kagome QSL candidate Cu$_3$Zn(OH)$_6$FBr. It does not experience any phase transition down to 50 mK, more than three orders lower than the antiferromagnetic Curie-Weiss temperature ($\sim$ 200 K). ${}^{19}$F NMR measurements reveal a gapped QSL ground state in Cu$_3$Zn(OH)$_6$FBr. The field dependence of the gap implies a zero-field gap $7.5\pm0.4$ K and spin-1/2 quantum number for  spin excitations, i.e. spinons.

We have successfully synthesized Cu$_3$Zn(OH)$_6$FBr polycrystalline samples by replacing the interkagome Cu$^{2+}$ sites in Cu$_4$(OH)$_6$FBr with non-magnetic Zn$^{2+}$. Our thermodynamical  (e.g. magnetic susceptibility and specific heat) measurements were carried out on the Physical Properties Measurement Systems (PPMS). The NMR spectra of $^{19}$F with the nuclear gyromagnetic ratio $\gamma=40.055$ MHz/T were obtained by integrating the spin echo as a function of the RF frequency at constant external magnetic fields of 0.914 T, 3 T, 5.026 T and 7.864 T, respectively.

Figure~\ref{fig:figure1} (a) and \ref{fig:figure1} (b) depict the crystal structure  of Cu$_3$Zn(OH)$_6$FBr. Micrometer-size crystals are easily observed by the scanning electron microscope (SEM) (Fig.~\ref{fig:figure1} (c)). The refinement of the powder X-ray diffraction pattern (Fig.~\ref{fig:figure1} (d)) shows that the material crystallizes in $P6_3/mmc$ space group with Cu$^{2+}$ ions forming a direct stack of undistorted kagome planes separated by non-magnetic Zn$^{2+}$ ions (Fig.~\ref{fig:figure1} (a) and (b)) as expected from theoretical calculations~\cite{Liu2015}. Cu$_3$Zn(OH)$_6$FBr is a charge-transfer insulator and the charge gap between Cu-3$d^9$ and O-2$p$ orbitals is around 1.8 eV according to first principles calculations~\cite{Liu2015a,Liu2015}. Powder X-ray diffraction measurements were carried out using Cu $K_{\alpha}$ radiation at room temperature. The diffraction data is analyzed by the Rietveld method using the program RIETAN-FP~\cite{Rietveld1969}. All positions are refined as fully occupied with the initial atomic positions taken from Cu$_4$(OH)$_6$FBr~\cite{Elliot2014}. The refined results are summarized in Table~\ref{tab:XRD}.

\begin{figure}
\centering
\includegraphics[width=\columnwidth]{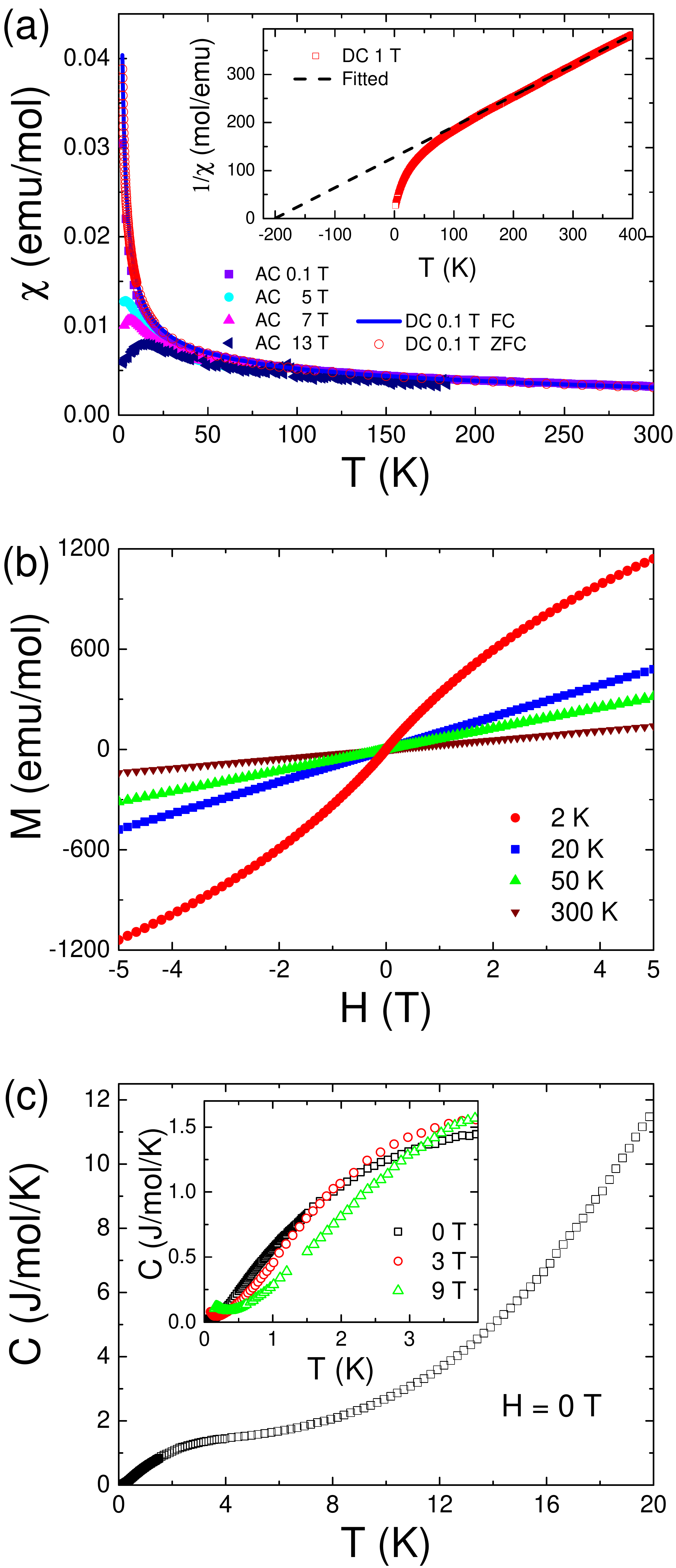}
\caption{(a) Temperature dependence of magnetic susceptibility under different magnetic fields measured by both DC and AC methods. In AC measurements, the oscillation filed amplitude is 17 Oe and the oscillation frequency is 633 Hz. The inset shows the temperature dependence of the inverse susceptibility 1/$\chi$ at 1 T. (b) Magnetic field dependence of magnetization at different temperatures. (c) Temperature dependence of the specific heat at zero field down to 50 mK. The inset shows the magnetic field effect on the specific heat at low temperatures.}
\label{fig:figure2}
\end{figure}
No phase transition is observed in our thermodynamical measurements (Fig.~\ref{fig:figure2}), establishing strong evidence for a QSL ground state in Cu$_{3}$Zn(OH)$_6$FBr. Temperature dependence of magnetic susceptibility under different magnetic fields does not display any magnetic transition down to 2 K as shown in Fig.~\ref{fig:figure2} (a). No splitting is detected between the field-cooled (FC) and zero-field-cooled (ZFC) results down to 2 K, indicating the absence of spin glass transition. At high temperature, magnetic susceptibility can be well fitted by the Curie-Weiss law  with the Curie temperature and Curie constant as -200 K and 1.57 K$\cdot$emu/mol, respectively. This indicates a strong antiferromagnetic superexchange interaction $J\sim17$ meV among Cu$^{2+}$ moments in the kagome planes. The $g$-factor is estimated to be about $g=2.4$, consistent with the $g$-factor measurements in the Barlowite~\cite{Han2016}. In Fig.~\ref{fig:figure2} (b),  no visible hysteresis loop is observed in the magnetic field dependence of magnetization at different temperatures. Figure~\ref{fig:figure2} (c) is the specific heat measurement at zero field down to 50 mK. The inset shows the magnetic field effect on the specific heat at low temperatures, which exhibits upturn behavior at high-field due to nuclear Schottky anomaly.


There are residual interkagome Cu$^{2+}$ (RIC) moments due to incomplete Zn$^{2+}$ substitution in Cu$_{3}$Zn(OH)$_6$FBr. Few Zn$^{2+}$ exists in kagome planes according to  the line shape of NMR spectra (see below in Fig.~\ref{fig:figure3}). The energy dispersive X-ray spectroscopy measurements at different locations indicate that the composition is stoichiometric with the atomic ratio between Cu and Zn as 1 : 0.36. The inductively coupled plasma atomic emission spectroscopy analysis suggests the atomic ratio between Cu and Zn as 1 : 0.30. From the chemical component analysis, we roughly estimate the concentration of the RIC moments to be $\sim10\%$, comparable to those in herbertsmithite~\cite{Freedman2010}. 

At low temperatures, RIC moments obscure the intrinsic kagome plane QSL behaviors in the bulk magnetic susceptibility and heat capacity, similar to previous results of herbertsmithite~\cite{Bert2007,Vries2008,Helton2010,Freedman2010,Han2016a,Kelly2016}. DC susceptibility at low temperatures in 0.1 T magnetic field is fitted by Curie-Weiss behavior with Curie constant and Curie temperature as 0.18 K$\cdot$emu/mol and -2.9 K, respectively, indicating weak antiferromagetically interacting RIC moments. Under high magnetic fields, the RIC moments freeze and the AC susceptibility drops at low temperatures (see Fig.~\ref{fig:figure2}. (a)). We also measure T-dependent AC susceptibilities for various frequencies and magnetic fields at low temperatures, see Fig.S~7 in the supplementary materials (SM)~\cite{suppl}. The AC susceptibility is independent of frequencies, implying that RIC moments do not develop spin glass freezing down to 2 K.   
The RIC moments also contribute a shoulder in the specific heat measurements at low temperatures (see Fig.~\ref{fig:figure2} (c)). The shoulder is supressed in magnetic fields, as shown in the inset of Fig.~\ref{fig:figure2} (c), along which the RIC moments are polarized, similar to herbertsmithite~\cite{Vries2008}. 

\begin{figure*}
\centering
\includegraphics[width=1.8\columnwidth]{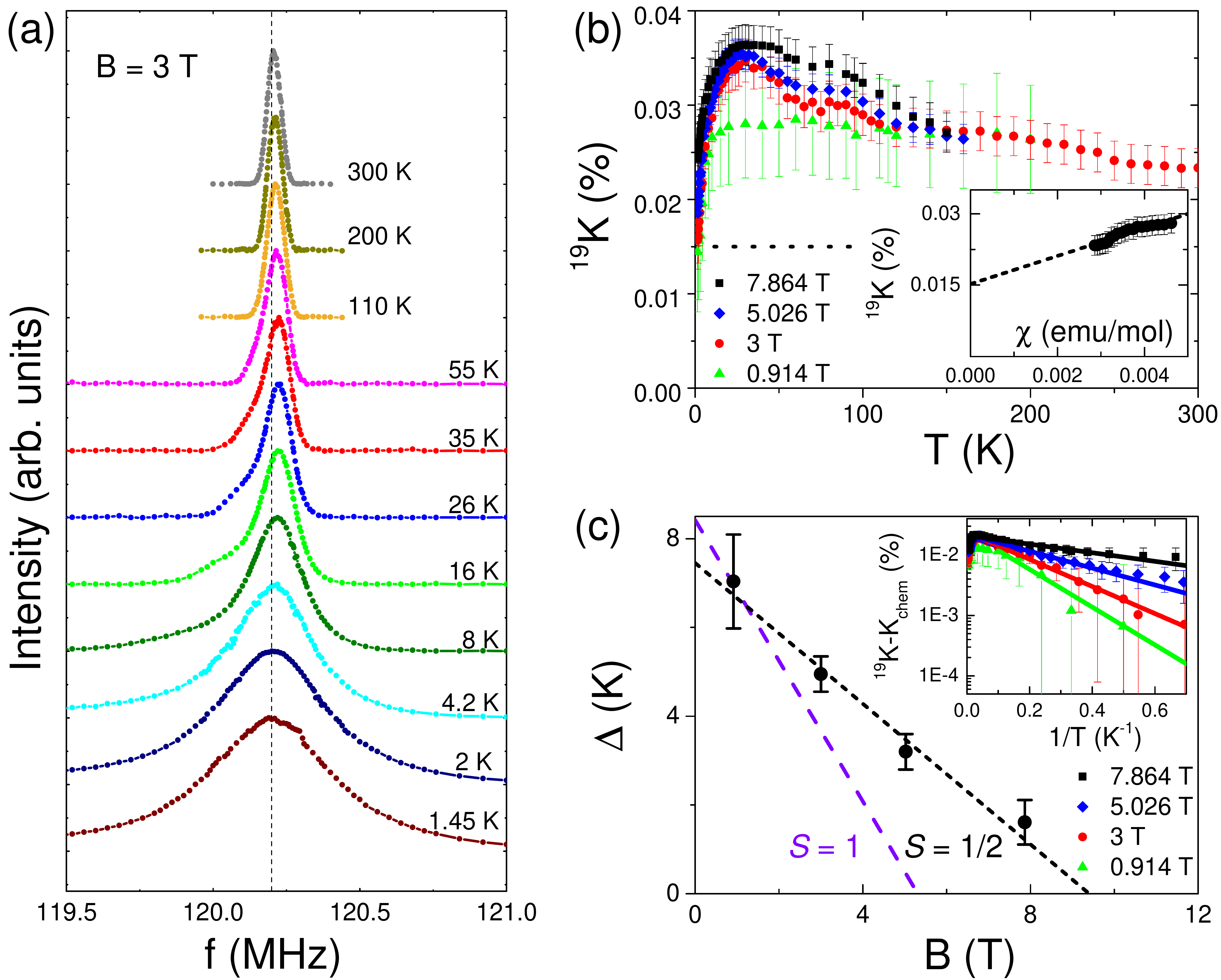}
\caption{(a) $^{19}$F NMR spectra under 3 T at different temperatures. 
The vertical dash line $f_0=120.199$~MHz, corresponding to the chemical shift, is a guide to the eyes. (b) Temperature dependence of the Knight shift $^{19}K$ determined from the peak positions of the spectra. The dotted horizontal line shows the position of $K_\text{chem}$ obtained from $^{19}K$-$\chi$ plot at high temperatures as shown in the inset. (c) Magnetic field dependence of the spin gap. The black short-dash line is fitted by $\Delta(B) = \Delta(0)-g\mu_\text{B} S B $ with spin quantum number $S = 1/2$. For comparison, we also plot $\Delta(B)$ for $S = 1$ shown by the blue dash line constrained by the value at 0.914 T, which hardly describes the data.  Inset shows the Arrhenius plot of $^{19}K-K_\text{chem}$ with the vertical axis in logarithmic scale, which demonstrates visually that the gap decreases with increasing magnetic field. The solid curve is the fitting function $A\exp(-\Delta/T)$ for $^{19}K-K_\text{chem}$.
}
\label{fig:figure3}
\end{figure*}

To directly unveil QSL physics in kagome plane, we implement NMR measurements to probe uniform spin susceptibility of kagome Cu$^{2+}$ spin moments in Cu$_3$Zn(OH)$_6$FBr. A unique advantage of Cu$_3$Zn(OH)$_6$FBr for the NMR measurements is that it contains $^{19}$F. 
It is known that $^{2}$D, $^{17}$O and $^{35}$Cl NMR measurements in herbertsmithite are rather difficult due to multiple resonance peaks resulted from nuclear spins $I=1$, $I=5/2$ and $I=3/2$, respectively~\cite{Imai2008,Olariu2008,Imai2011,Fu2015}. In contrast, only one resonance peak needs to be resolved for $^{19}$F with $I=1/2$ nuclear spin, as shown in Fig.~\ref{fig:figure3} (a). The sharp high-temperature peaks suggest that few Zn$^{2+}$ exists in kagome planes. Moreover, no extra peak due to RIC moments is observed even at low temperatures. The line shape asymmetry may arise from the magnetic anisotropy, e.g. $g_\|/g_\perp=2.42/2.21$ in Barlowite~\cite{Han2016}. We have also carried out the measurements with different pulse interval ($\tau$) in NMR echo to exclude the possibility of impurity moment contributions in the NMR spectrum~\cite{suppl}.

In a gapped QSL, the spin susceptibility should become zero at low temperature. The Knight shift is related to the uniform susceptibility $\chi$ as $^{19}K = A_\text{hf}\chi + K_\text{chem}$, where $A_\text{hf}$ is the hyperfine coupling constant between the $^{19}$F nuclear spin and the electron spins and $K_\text{chem}$ is  the $T$-independent chemical shift. $K_\text{chem} = 0.015\%$ is obtained from $^{19}K$-$\chi$ plot at high temperatures as shown in the inset of Fig.~\ref{fig:figure3} (b), where $\chi$ is DC susceptibility at $B=3$ T. Figure~\ref{fig:figure3} (b) shows that the Knight shift drops quickly below $\sim30$ K. At high temperatures ($\sim 100$ K), 
Knight shift $^{19}K$ has a systematic variation as a function of magnetic field, whose origin is unclear at present and left for future investigation, but we note that such a behavior would not change our results at low temperatures below 30 K. 
The Knight shift at low fields (0.914 T and 3 T) tends to merge to $K_\text{chem}$ at low temperatures, similar to previous results of herbertsmithite~\cite{Fu2015}. The inset of Fig.~\ref{fig:figure3} (c) is the Arrhenius plot of ${}^{19}K-K_\text{chem}$, where the low-temperature data can be well fitted by an exponential function $A\exp(-\Delta/T)$, with $A$ and $\Delta$ as fitting parameters for a constant and the gap value, respectively. In the fit, we fixed $K_\text{chem} = 0.015\%$. 

With elevating the magnetic fields, the gap is suppressed due to Zeeman effect as $\Delta(B) = \Delta(0)-g\mu_\text{B} S B $, where $\mu_\text{B}$ is Bohr magneton. From the linear fitting of the field dependence of $\Delta$, we obtain a zero-field gap $7.5\pm0.4$~K and $gS=1.16\pm0.11$. Regarding to $g = 2.4$ obtained from bulk magnetic susceptibility measurements in Fig.~\ref{fig:figure2} (a), $gS=1.16\pm0.11$ confirms a spin quantum number $S=1/2$ and $g=2.32\pm0.22$.  
The spin $S=1/2$ quantum number implies fractionalized spinon excitations in the quantum spin liquid compound  Cu$_3$Zn(OH)$_6$FBr.

Detecting spin-1/2 quantum number of spin excitations in a QSL state is of great significance. Spin-1/2 spinon excitations have been discussed since the early stage of spin liquid theory~\cite{Kivelson1987}, yet there is no direct experimental confirmation of the spin-1/2 quantum number till now. Our results show that Cu$_3$Zn(OH)$_6$FBr has a gapped QSL ground state, consistent with results in herbertsmithite~\cite{Fu2015} and unambiguously manifest the spin-1/2 quantum number of spinons. It reflects the spin fractionalization in a QSL state when spin rotation symmetry meets topology. Within minimal symmetry (e.g. time reversal symmetry and translational symmetry) assumptions, a gapped kagome QSL should be $\mathbb{Z}_2$-gauge type~\cite{Read1991,Wen1991} (i.e. toric code type~\cite{Kitaev2003}) according to the theoretical constraints~\cite{Zaletel2015}. 

In conclusion, we have successfully synthesized a new kagome compound Cu$_3$Zn(OH)$_6$FBr and its quantum spin liquid ground state is verified in our thermodynamical measurements. Our $^{19}$F NMR data reveals a gapped spin-liquid ground state for Cu$_3$Zn(OH)$_6$FBr, similar to previous $^{17}$O NMR results on herbertsmithite. Most importantly, we provide experimental evidence for spin-1/2 quantum number for spin excitations, i.e. spinons. We therefore believe that Cu$_3$Zn(OH)$_6$FBr provides a promising platform for future investigations of the topological properties of quantum spin liquid states. 

We acknowledge Yongqing Li for discussions on the magnetic susceptibility measurements. We thank Xi Dai and Zhong Fang for useful discussions. 
We acknowledge fundings from the National  Key  Research and  Development Program of China under Grant Nos. 2016YFA0300502, 2016YFA0300503, 2016YFA0300604, 2016YF0300300 and 2016YFA0300802, the National Natural Science Foundation of China under Grant Nos. 11421092, 11474330, 11574359, 11674406, 11374346 and 11674375, National Basic Research Program of China (973 Program) No. 2015CB921304, the National Thousand-Young-Talents Program of China, the Strategic Priority Research Program (B) of the Chinese Academy of Sciences  under Grant No. XDB07020000, XDB07020200 and XDB07020300. The work in Utah is supported by DOE-BES under No. DE-FG02-04ER46148.

\bibliography{ZnBarlowite}

\end{document}